\documentclass[pre,onecolumn,a4paper,12pt]{revtex4}

\usepackage{amsmath}
\usepackage{graphicx}

\newcommand{\Cee}{\mathcal{C}}
\newcommand{\X}{\mathcal{X}}

\begin{document}

\title{Reversibility, heat dissipation and the importance of the thermal environment in stochastic models of nonequilibrium steady states}

\author{R.\ A.\ Blythe} \affiliation{SUPA, School of Physics,
University of Edinburgh, Mayfield Road, Edinburgh EH9 3JZ, UK}

\date{\today}

\begin{abstract}
We examine stochastic processes that are used to model nonequilibrium processes (e.g, pulling RNA or dragging colloids) and so deliberately violate detailed balance. We argue that by combining an information-theoretic measure of irreversibility with nonequilibrium work theorems, the thermal physics implied by abstract dynamics can be determined. This measure is bounded above by thermodynamic entropy production and so may quantify how well a stochastic dynamics models reality. We also use our findings to critique various modeling approaches and notions arising in steady-state thermodynamics.
\end{abstract}


\maketitle

A theory of nonequilibrium physics is vital if we are to understand such diverse phenomena as geological or biological processes which are inherently dissipative in nature.  Although a general theory remains both challenging and elusive, it is now possible to obtain precise experimental data for mesoscopic objects such as RNA strands \cite{LDSTB02} and optically-trapped colloids \cite{WSMSE02} undergoing irreversible manipulation.  In turn this has allowed  theoretical developments, such as strikingly general \emph{nonequilibrium work relations}, to be verified \cite{BLR05}.

In this work, we address the fundamental question of how to faithfully model irreversible, dissipative physics with stochastic dynamics.  We introduce an irreversibility measure for stochastic processes which, in contrast to standard expressions, respects such basic physics as frame invariance.  We find that an explicit prescription for a system's thermal environment---often absent in models---is essential if predictions for entropy production are even to be possible.  Using work relations for stochastic systems \cite{HS07} we find our main result, inequality (\ref{dIS}) below, which shows that such predictions always underestimate the true dissipation, unless all relevant processes are modeled.  This suggests that a model predicting less dissipation than is observed is incomplete; and one predicting more should be rejected. Since our results hold for arbitrary nonequilibrium states, we gain many insights into theories and models of \emph{nonequilibrium steady states} (NESS) \cite{Reviews} which cannot not be drawn from, for example, a similar expression recently derived for isolated systems constrained initially to be at equilibrium \cite{KPB07}.

We begin by reviewing the modeling paradigm introduced by 
Katz, Lebowitz and Spohn in their seminal work on fast ionic conductors \cite{KLS83}.  One takes the master equation for a (discrete-time) process, $P_{t+1}(\Cee) = \sum_{\Cee'} P_t(\Cee') M(\Cee'\to\Cee)$ in which $P_t(\Cee)$ is the time-dependent distribution of microstates $\Cee$.  For a reversible, equilibrium system, the transition probabilities $M(\Cee\to\Cee')$ are taken to satisfy the detailed balance condition
\begin{equation}
\label{db}
P^\ast(\Cee) M(\Cee\to\Cee') = P^\ast(\Cee') M(\Cee'\to\Cee)
\end{equation}
with respect to the Boltzmann distribution $P^\ast(\Cee)={\rm e}^{-\beta E(\Cee)}$ where $\beta$ is inverse temperature and $E$ the internal energy.  This relation guarantees \emph{microscopic reversibility} \cite{Kelly79,Books} in the steady state---i.e., that any sequence of configurations is witnessed with the same probability as its time reversal.  To model irreversible physics, and in particular a NESS, one must deliberately violate detailed balance.

\begin{figure}
\includegraphics[width=0.66\linewidth]{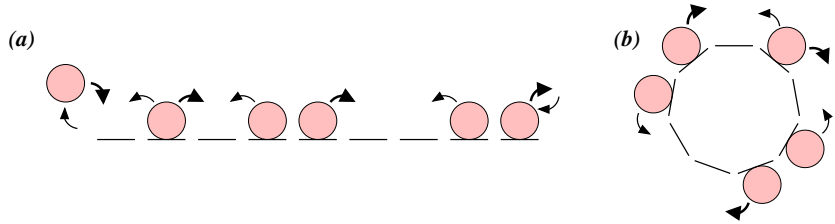}
\caption{\label{aseps} Biased diffusion arising from the local/generalized detailed balance principles applied to hard-core particles in a one-dimensional linear potential gradient under (a) open and (b) periodic boundary conditions.}
\end{figure}

There is no obviously correct way to go about this, so following \cite{KLS83} it is commonplace to invoke a \emph{local} (or \emph{generalized} \cite{BD07}) detailed balance principle.  In this approach, (\ref{db}) is taken to apply over some closed subset of configurations, and a nonequilibrium system formed by joining together subsystems that are in contact with heat baths at \emph{different} temperatures.  For illustrative purposes, we take the specific example of hard-core particles in a one-dimensional linear potential, which if connected to particle reservoirs at different densities, would exhibit the biased diffusion shown in Fig.~\ref{aseps}a.  Alternatively, periodic boundary conditions might be imposed (Fig.~\ref{aseps}b), at the expense of being able to couch the dynamics in terms of a single-valued potential.  Note, however, this modeling procedure can be used for all types of particle interaction and in any dimension.

A problem with this approach is that one loses sight of how the system interacts thermally and mechanically with its environment, and could thus be argued to lack a firm physical basis.  Furthermore, it is not obvious that alternative approaches, e.g., those based on maximal entropy analyses subject to macroscopic flux constraints \cite{Dewar03,rmle}, offer more realistic descriptions of nonequilibrium physics than the model-building tradition described above.  We address these shortcomings by introducing a framework in which a model system's thermal environment is made completely explicit, which, as we now show, is necessary to establish the degree to which a stochastic dynamics is irreversible.

The standard way to do this is to compare the left- and right-hand sides of the detailed balance condition (\ref{db}).  For example, logarithms of their ratio appear in a widely-used expression for the entropy production attributed to Schnackenberg \cite{Schnackenberg76}, an action functional that exhibits a Gallavotti-Cohen symmetry \cite{LS99} and various time-dependent generalizations \cite{Seifert05,HS07}, as well as the \emph{housekeeping heat} \cite{OP98} that is instantaneously dissipated by a NESS \cite{Housekeeping}.  Their differences, meanwhile, have been proposed to characterize a NESS \cite{SZ06}, since instantaneous physical currents (which vanish at equilibrium) can be derived from them.  The violation of (\ref{db}) is thus almost universally used to recognize a dynamics with a dissipative steady state, despite the obvious shortcoming that an observer can witness the former without the latter simply by changing frame.

\begin{figure}
\includegraphics[width=0.66\linewidth]{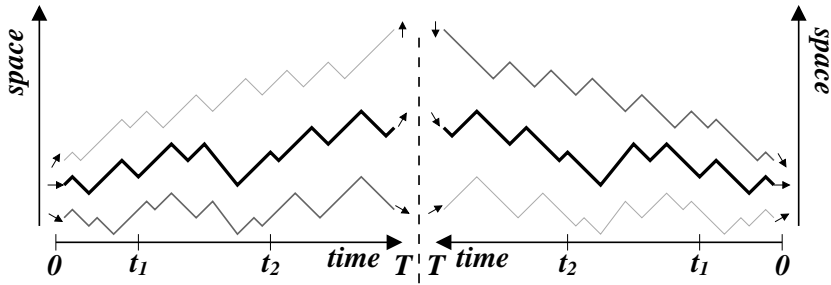}
\caption{\label{trev} Comparison of trajectories generated by the forward and reverse processes.  Forward trajectories of length $T$ are generated, at which point all velocities (shown as short arrows) are flipped, and the reverse dynamics started.  The heaviness of the lines indicates the probability of the trajectories in each ensemble.  The central trajectory appears with the same probability in both ensembles, whereas the outer trajectories appear with different probabilities, thus indicating an irreversible dynamics.}
\end{figure}

This difficulty is resolved by realizing that when comparing the probability of a trajectory with that of its time reversal, the latter should not be drawn from the same ensemble as the former, but from an ensemble in which all degrees of freedom in the \emph{environment} are \emph{also} time reversed.  The dynamics that generate this second ensemble we shall call the \emph{reverse process}.  We may now define the following general measure of reversibility for any stochastic dynamics, i.e., not restricting ourselves to ergodic time-homogeneous Markov chains in discrete time (see also Fig.~\ref{trev}).  Let $\X$ denote a trajectory $(\Cee_1, \Cee_2, \ldots, \Cee_n)$ that visits configuration $\Cee_i$ at time $t_i$, possibly other (unspecified) configurations at other times and eventually reaches, with probability $P_T(\Cee_T)$, configuration $\Cee_T$ at a time $T>t_n$.  Given an initial configuration $\Cee_0$ that is drawn from a distribution $P_0(\Cee_0)$, this trajectory is taken to appear under the forward dynamics with probability $P(\X|\Cee_0)$.  This is to be compared with the probability of seeing the time-reversed trajectory $\hat{\X}$, in which the image $\hat{\Cee}_i$ of $\Cee_i$ under time reversal (i.e., with all velocities reversed) is seen at time $t_i$ running backwards from time $T$ to $0$.  Given a starting configuration $\hat{\Cee}_T$ drawn from a distribution $\hat{P}_T(\hat{\Cee}_T)$, this reverse trajectory appears with probability $\hat{P}(\hat{\X}|\hat{\Cee}_T)$.  If there is to be any possibility for the ensembles of forward and reverse trajectories to coincide, we must take $\hat{P}_T(\hat{\Cee}_T) = P_T(\Cee_T)$, i.e., start the reverse process by immediate time reversal of configurations reached after time $T$ under the forward dynamics.  Any other choice requires us to make additional assumptions on the dynamics.

In the spirit of Landauer's principle \cite{Landauer61}, we now loosely associate information lost under the dynamics---quantified here by the additional information required to reconstruct the forward trajectory ensemble from the reverse---with irreversibility and dissipation.  This amount of information (in natural units) is given by \emph{relative entropy} of the two ensembles \cite{CT06},
\begin{equation}
\label{dI}
\Delta I = 
\sum_{\Cee_0,\X,\Cee_T} P_0(\Cee_0) P(\X|\Cee_0) \ln \frac{P_0(\Cee_0) P(\X|\Cee_0)}{P_T(\Cee_T) \hat{P}(\hat{\X}|\hat{\Cee}_T)} \;.
\end{equation}
To make contact with thermal physics, we assume that just before the start of the forward and reverse processes, any heat baths present are manipulated by a thermostat in such a way that the probability that any particular bath configuration is realized is given by the Boltzmann distribution with a well-defined temperature.  Note that this necessarily requires correlations between the system and bath to vanish rapidly---this is the origin of dissipation, as will be seen concretely below.  Under such conditions, Jarzynski's detailed fluctuation relation \cite{Jarzynski00}
\begin{equation}
\label{jar}
\ln \frac{P(\X, \Delta S_{\rm env} | \Cee_0 )}{\hat{P}(\hat{\X}, -\Delta S_{\rm env} | \hat{\Cee}_T )} = \Delta S_{\rm env}
\end{equation}
applies (in a system of units where Boltzmann's constant is unity).
Here, $\Delta S_{\rm env}$ is the total entropy increase in the heat baths under the forward dynamics.  This is a random variable if the trajectory $\X$ contains insufficient detail to determine how much energy has been exchanged with each heat bath separately.  In deriving this formula, it was assumed that the microscopic evolution is Hamiltonian with respect to potentials that are time-independent in the baths but may exhibit time dependence in the system of interest.  Stochasticity enters from the Boltzmann sampling of bath configurations and any coarse-graining in the specification of the trajectory $\X$.

We finally arrive at an important inequality---the main result of this work---by averaging over $\Delta S_{\rm env}$.  An application of the log-sum inequality \[ \sum_{i} a_i \ln (a_i/b_i) \ge \sum_i a_i \ln ( \sum_i a_i / \sum_i b_i) \] (which itself is a consequence of Jensen's inequality \cite{CT06}) leads to
\begin{equation}
\label{dIS}
0 \le \Delta I \le \langle \Delta S_{\rm env} \rangle + S_G(T) - S_G(0)
\end{equation}
in which $S_G(t)= -\sum_{\Cee} P_t(\Cee) \ln P_t(\Cee)$, the Gibbs entropy of the distribution at time $t$ under the forward dynamics. Whilst a similar result was recently given for isolated systems starting at equilibrium \cite{KPB07}, our result (\ref{dIS}) holds for \emph{any} initial condition and explicitly requires the system to be open to the environment. Moreover, the thermostatting of the baths means that $S_{\rm env}$ is the true entropy production, which is not always true of isolated systems \cite{PR07}. As we now discuss, (\ref{dIS}) thus provides hitherto unavailable information---spatial and temporal---about heat production in a general nonequilibrium system, e.g., a NESS.

For example, the lower bound is attained only if every forward trajectory appears with the same probability as its time-reversal in the reverse ensemble (i.e., no information loss occurs and the process is reversible). This leads to an extended detailed balance relation for a NESS, viz,
\begin{equation}
\label{mdb}
P^\ast(\Cee) M(\Cee\to\Cee') = P^\ast(\Cee') \hat{M}(\hat{\Cee}'\to\hat{\Cee}) \;.
\end{equation}
Note that one cannot decide on the reversibility of a dynamics until its reversal $\hat{M}(\hat{\Cee}'\to\hat{\Cee})$ has been identified (see below for concrete examples).  Since equality of forward and reverse trajectory sets implies $P^\ast(\Cee) = \hat{P}^\ast(\hat{\Cee})$, one finds (\ref{mdb}) can be written in a more symmetric form with $\hat{P}^\ast(\hat{\Cee}')$ on the right-hand side.  This condition is equivalent to (\ref{mdb}), as can be shown from conservation of probability $\sum_{\Cee'} M(\Cee\to\Cee')=1$.  The condition (\ref{mdb}) can also be stated as a Kolmogorov criterion \cite{Kelly79}
\begin{equation}
\label{kc}
M(\Cee_1\to\Cee_2) M(\Cee_2\to\Cee_3) \cdots M(\Cee_{T}\to\Cee_1) =
\hat{M}(\hat{\Cee}_1\to\hat{\Cee}_T) \hat{M}(\hat{\Cee}_T\to\hat{\Cee}_{T-1}) \cdots \hat{M}(\hat{\Cee}_{2}\to\hat{\Cee}_1)
\end{equation}
on every loop in configuration space of length $T \ge 1$.  This allows reversibility to be decided without prior knowledge of the stationary distributions $P^\ast(\Cee)$ or $\hat{P}^\ast(\hat{\Cee})$.  Equivalence of (\ref{kc}) and (\ref{mdb}) is shown in a similar way to the standard case \cite{Kelly79}.

The upper bound in (\ref{dIS}) is reached only if the stochastic dynamics faithfully models all dissipative processes in the physical system.  This we have already seen from the fact that if one cannot work out from the trajectory $\X$ how much energy has been exchanged with each bath, $\Delta S_{\rm env}$ in (\ref{jar}) is a random variable and $\Delta I$ underestimates the true entropy change. As for isolated systems starting at equilibrium \cite{KPB07}, the log-sum inequility implies that $\Delta I$ further decreases under spatial coarse-graining. Since in the present, more general context, $\Delta I$ contains temporal information, coarse-graining in time, or reduction of a non-Markov dynamics to a Markov process, has the same effect.  We thus suggest that this reduction in $\Delta I$ could reveal the amount of heat dissipated at the finer-grained scale, and that differences between a model's prediction for entropy production and that measured in a real system might allow deficiencies in the model to be identified.

Finally, we use relation (\ref{dIS}) to gain new insights into the stochastic modeling approaches described earlier.  The physics of the open system (constructed using generalized detailed balance \cite{BD07}, Fig.~\ref{aseps}a) is consistent with that described above.  In particular, the interpretation of $\ln M(\Cee\to\Cee')/M(\Cee'\to\Cee)$ as being proportional to the energy exchanged with a reservoir holds, as long as one is confident that all dissipative processes are captured by the Markov dynamics of particles hopping on a lattice, and further that one can unambiguously identify which bath exchanges energy at any given transition in the system of interest.  Note that since particle velocities are not included in the model $\hat{\Cee} \equiv \Cee$; also $\hat{M}\equiv M$ as the potential is time-independent.  We also see explicitly that dissipation results from a continuous thermostatting of the reservoirs that enables particles to enter or leave the system with a constant probability in every timestep.

Models in which a current is induced by periodic boundary conditions (see Fig.~\ref{aseps}b) are more subtle.  There are at least two ways in which such dynamics may be realized in a manner consistent with (\ref{jar}).  First, one can apply a change of frame to unbiased diffusion on a ring, and then discretize: clearly, this yields a reversible dynamics.  Alternatively, one can fashion a time-homogeneous Markov process by coarse-graining the response to a rotating potential over one period of its motion.  For concreteness, and to keep track of all energy fluxes, we consider a dynamics in which the energy function $E_t(\Cee)$ is static during each timestep, and changed instantaneously between them.  As in \cite{Crooks99}, the dynamics is assumed to satisfy (\ref{db}) whilst the potential is static.  One can compute transition probabilities over the course of one period of rotation of a potential (e.g., a square well) in either direction, and show that typically the forward and reverse dynamics, $M$ and $\hat{M}$, are not simply related to each other, nor do they satisfy (\ref{kc}) \cite{unpub}.  Due to the coarse-graining, the irreversibility measured by $\Delta I$ underestimates the true dissipation in the system.

We remark that in our framework, coarse-graining generically leads to (and is in fact the only mechanism for) the appearance of nonconservative forcing in the system of interest.  By contrast, such forces are central to models based on Langevin equations (see, e.g., \cite{Kurchan98,Housekeeping,Seifert05}), are associated with the dissipation of housekeeping heat \cite{OP98} and have been argued to differ fundamentally from those due to a moving potential.  Since coarse-graining blurs this distinction, it is not clear in what sense it is meaningful.  We also remark that the nonconservative forces considered in \cite{Housekeeping,Seifert05,HS07} are assumed not to change under time reversal, which even in simple models is not the case for forces due to a moving potential.  As well as this, it seems often to have been assumed that trajectories are sufficiently detailed that the upper bound in (\ref{dIS}) is in fact an equality, and further that housekeeping heat can be defined on a per-trajectory basis in terms of the instantaneous state of a system and its environment \cite{Housekeeping,HS07}.  Such a definition conflicts with the macroscopic quantity described in \cite{OP98} if the latter is interpreted as the heat exported by some sequence of dissipative steady states if one could somehow switch between them without incurring additional entropy costs.  For these costs to be removed when averaging over all microscopic realizations of an arbitrary switching process, one finds that details of the history of this process must appear in the single-trajectory expressions, contrary to \cite{Housekeeping,HS07}.  We thus contend that far greater clarity about the meaning of central quantities in the putative framework of steady-state thermodynamics \cite{OP98} is necessary.

Finally, we examine modeling approaches in which transition probabilities are obtained from maximal-entropy inference subject to macroscopic flux constraints \cite{rmle}.  If this is to be interpreted as a general recipe for deriving a stochastic dynamics, then we have shown the need to derive both the forward and the reverse dynamics, the latter obtained from time-reversal of all driving forces, using this procedure.  If all macroscopic fluxes simply change sign under time reversal, the outcome will be a reversible dynamics, and so---at least within the framework put forward here---one needs to argue for time-asymmetric macroscopic constraints to realize a dissipative dynamics.  However, the theory developed in \cite{rmle} is intended to apply to internal portions of a larger sheared system, and as such are in contact with \emph{nonequilibrium} reservoirs, not the thermostatted heat baths described here.  It would be interesting to try and interpret our definition of reversibility in this more general context. 

In summary, we have argued, by examining what it means for a stochastic process to be reversible, that the presence of dissipation in a model steady state can only be decided once a reverse process, which demands knowledge of the environment, is known.  Using Jarzynski's detailed fluctuation theorem (\ref{jar}) and results from information theory, we have specified a physical environment that allows information loss to be bounded above by the thermodynamic entropy production, extending to a much larger class of nonequilibrium systems a result of \cite{KPB07}.  This allows physical mechanisms by which the system is driven and heat dissipated away to be identified in otherwise abstract models of a NESS, illustrating with the particular examples shown in Fig.~\ref{aseps}.  We note that although the standard detailed balance condition (\ref{db}) is satisfied in all these models, only in some is the steady state actually dissipative.  Although we have couched our discussion in terms of discrete-time Markov processes, everything we have said also applies in the continuous-time limit.

Whilst we have mostly taken a theoretical perspective, we hope that the main result (\ref{dIS}) will be useful experimentally, e.g., to determine whether a stochastic model captures all relevant dissipative processes, as we have proposed.  The hypothesis that decreases in $\Delta I$ under coarse-graining relate to dissipation at a given scale could also be tested explicitly.  Finally, we see, from the difficulty in discriminating between nonconservative forces and those due to coarse-graining a moving potential for example, that in the field of nonequilibrium statistical mechanics conceptual problems remain.

I thank Mike Evans, Andy Jackson and Wilson Poon for useful comments and the Royal Society of Edinburgh for the award of a Personal Research Fellowship.

\end{document}